# Big consequences of small changes
## (Non-locality and non-linearity of Hartree-Fock equations)

M. Ya. Amusia[1][1,2]

[1] *Racah Institute of physics, The Hebrew University, 91904 Jerusalem, Israel*
[2] *Ioffe Physico-Technical Institute,194021 St. Petersburg, Russia*

**Key words** Hartree-Fock equation, wave-functions asymptotic, Levinson's theorem, Green's function

**PACS** 03.65.Ca, 03.65.Ge, 03.65.Nk, 03.65.Xp

**Abstract**

It is demonstrated that non-locality and non-linearity of Hartree-Fock equations dramatically affect the properties of their solutions that essentially differ from solutions of Schrödinger equation with a local potential. Namely, it acquires extra zeroes, has different coordinate asymptotic, violates so-called gauge-invariance, has different scattering phases at zero energy, has in some cases several solutions with the same set of quantum numbers, usually equivalent expressions of current and Green's functions became non-equivalent. These features result in a number of consequences for probabilities of some physical processes, leading e. g. to extra width of atomic Giant resonances and enhance considerably the ionization probability of inner atomic electrons by a strong field.

## 1. Introduction

There is no doubt that the surrounding us macroscopic world being the subject of chemistry, botanic and biology can in principle be described by a huge Schrödinger equation that include all constituent particles – atomic nuclei and electrons – their kinetic energy and interaction. However, such an equation is virtually useless even for a many-electron atom. In reality, many mean any number $N \geq 3$. In 1928 D. Hartree discovered equations [1] that have dramatically simplified the very complicated problem of solving Schrödinger equation for N particles reducing it to a system of N one – particle equations, which describe independent motion of all the system N constituents, moving in the so-called *self – consistent* field, created by common action of all but the considered particle. Apparently, Hartree was the first who introduced non-linearity into one-particle Schrödinger equation. His functions have evident minuses – for different states they are in general non-orthogonal and the total wave-function of the system has no definite symmetry.

Both minuses were eliminated by V. A. Fock [2], who in accord with general requirements added to the Hartree idea of the self-consistent field the anti-symmetrization of the total wave-function, thus accounting for the fact that identical particles are indistinguishable. This lead to additional so-called exchange *Fock* term in the equation now called Hartree-Fock (HF) equation, which is non-local. HF equation are one of the most essential and powerful in quantum physics, from nuclear to atomic, molecular, solid, gas and liquid state physics.

While HF equations are extensively used, some specific features of them that result from non-linearity and non-locality were not explicitly analyzed and known to physics community. It is the aim of this paper to eliminate this defect. The Fock term is much smaller than the Hartree term. But the consequences of its introduction are very essential. Here we present some profound consequences resulting from non-locality and non-linearity of Hartree-Fock (HF) equations that describe a whole variety of many-particle quantum systems amazingly accurate already in one-particle approximation. Note that textbooks on quantum mechanics, even advanced, limit themselves with local and linear one-particle Schrödinger equation. This is why we will show the difference that comes from non-linearity and non-locality of the potential.

---

[1] E-mail: amusia@vms.huji.ac.il

As a concrete example of a many-particle system we will consider atoms with N electrons, $N \gg 1$. As inter-particle interaction we consider only pair forces that depend only upon the inter-particle distance. Generalization to spin-dependent forces is straightforward. Velocity-dependent Breit -type interaction [3] The nuclear mass will high accuracy can be considered infinitely big in comparison to the total mass of electrons. In reality, the respective mass ratio is more than 4000.

**2. Schrödinger equation**

It is well known that an atom with a two-particle pair interaction between its N electrons is described by the following Schrödinger equation[2]

$$\left[\sum_{i=1}^{N}\left(-\frac{\Delta_i}{2} + U(r_i)\right) + \frac{1}{2}\sum_{i \neq j=1}^{N} V(r_{ij})\right]\Psi(\vec{r}_1...\vec{r}_N) \equiv \hat{H}_A \Psi(\vec{r}_1...\vec{r}_N) = E\Psi(\vec{r}_1...\vec{r}_N). \quad (1)$$

The electron density $\rho(\vec{r}_1...\vec{r}_N)$ is given by the relation $\rho(\vec{r}_1...\vec{r}_N) = |\Psi(\vec{r}_1...\vec{r}_N)|^2$ and the one-electron density is given by

$$\rho(r) = \int |\Psi(\vec{r}...\vec{r}_N)|^2 \, d\vec{r}_2 \cdots d\vec{r}_N. \quad (2)$$

We consider $U(r_i) = -Z/r_i$ and $V(r_{ij}) = 1/|\vec{r}_i - \vec{r}_j|$, where Z is the nuclear charge. The atomic Hamiltonian $\hat{H}_A$ is local and linear. For such a Hamiltonian particles velocity have two identical expressions [4]

$$\vec{v} \equiv \partial \vec{r} / \partial t = i[\hat{H}_A, \vec{r}] = i\hat{\nabla}. \quad (3)$$

This leads to two identical expressions for electric current that in atomic system of units coincides with expressions (3). The operator of photon-electron interaction is given by the following expression [3]

$$V_{pe} = \frac{1}{c}\vec{A}(\vec{r},t)\vec{v}, \quad (4)$$

where $\vec{A}(\vec{r},t)$ is the electromagnetic field vector potential and c is the speed of light.

Another peculiar feature of the precise Hamiltonian $\hat{H}_A$ is that the dipole oscillator strengths $f_v$ and the dipole photoionization cross section $\sigma(\omega)$ satisfy the following sum rule

---

[2] Atomic system of units is used in this paper: the electron mass, charge and Planck constant are equal to 1: $m = e = \hbar = 1$

[5]

$$\sum_v f_v + \frac{c}{2\pi^2}\int_I^\infty \sigma(\omega)d\omega = N, \quad (5)$$

where $v$ denote the discrete excited atomic state and $I$ stands for the atomic ionization potential – minimal photon energy required the electron to move off the atom.

For the discussions in this paper it is essential to have in mind that for approximate Hamiltonians equations (3) and (5) are violated.

## 3. Hartree-Fock (HF) equation

The idea of Hartree was to approximate $\varphi_j(\vec{r})$ in (1) as a product of so-called one – particle wave-functions $\varphi_j(\vec{r})$: $\Psi(x_1...x_N) \approx \Psi_H(x_1...x_N) = \prod_{1\leq j \leq N} \varphi_j(x_j)$, where $x \equiv \vec{r}, s$ is the combination of the radial $\vec{r}$ and spin $s$ coordinate of the electrons. Demanding a minimum of the expression $\int \Psi_H^*(x_1...x_N) \hat{H}_A \Psi_H(x_1...x_N) \prod_{1\leq j \leq N} dx_j$ [^3] relative to variation of $\varphi_j(x)$ functions, one obtains the following equations:

$$-\frac{\Delta}{2}\varphi_j(x) - \frac{Z}{r}\varphi_j(x) + \sum_{k=1}^N \int \varphi_k^*(x') \frac{1}{|\vec{r}'-\vec{r}|}\varphi_k(x')\varphi_j(x)dx' \equiv$$
$$\equiv -\frac{\Delta}{2}\varphi_j(x) + U_H(r)\varphi_j(x) = E_j \varphi_j(x) \quad (6)$$

The functions $\varphi_j(\vec{r})$ and $\varphi_k(\vec{r})$ for $j \neq k$ are orthogonal. The asymptotic of $U_H(r)$ is $(-Z+N)/r$ For neutral atoms the potential $U_H(r)$ is of short range and have at most one or two discrete excited levels, the wave functions of which is determined by (6) with $j$ being a level above N already occupied. This is in direct contradiction to what is known about real atoms.

Introducing the Hartree electron density $\rho_H(\vec{r}) = \sum_{k=1}^N \int \varphi_k^*(\vec{r})\varphi_k(\vec{x})$, (6) transforms into

$$\left[-\frac{\Delta}{2} - \frac{Z}{r} + \int d\vec{r}' \frac{1}{|\vec{r}'-\vec{r}|}\rho_H(\vec{r}')\right]\varphi_j(x) \equiv -\frac{\Delta}{2}\varphi_j(x) + U_H(r)\varphi_j(x) = E_j \varphi_j(x). \quad (7)$$

It is obvious that (7) includes an unphysical effect: the action of electron $j$ upon itself since $\rho_H(\vec{r})$ is the total electron density of an atom in approximation (6). Hartree himself modified these equations eliminating self action artificially, substituting $\rho_H(\vec{r})$ by $\rho_{H(j)}(\vec{r}) \equiv \rho_H(\vec{r}) - |\tilde{\varphi}_j(\vec{r})|^2$. With this substitution $U_H(r)$ asymptotic became $(-Z+N-1)/r$ that is $-1/r$ even for a neutral atom. Such a potential supports an infinite number of discrete excited levels, the wave functions of which is determined by (7) with $j$ being a level above N already occupied.. However, functions $\tilde{\varphi}_j(\vec{r})$ and $\tilde{\varphi}_k(\vec{r})$ became non-

---
[^3]: Integration over spin variable means summation over two electron spin projections.

orthogonal for $j \neq k$. The second defect of Hartree approximation is that $\Psi_H(x_1...x_N)$ for atomic electrons do not satisfy Pauli principle: when coordinates of two electrons coincide $x_j = x_k$ this wave function is not equal to zero.

Both minuses were eliminated by Fock in [2], when instead of $\Psi_H(x_1...x_N)$ he have introduced $\Psi_{HF}(x_1...x_N) = \hat{A} \prod_{1 \leq j \leq N} \varphi_j(x_j)$, where $\hat{A}$ is the operator of anti-symmetrization over all coordinates. Then again demanding a minimum of the expression $\int \Psi^*_{HF}(x_1...x_N) \hat{H}_A \Psi_{HF}(x_1...x_N) \prod_{1 \leq j \leq N} dx_j$ relative to variation of $\varphi_j(x)$ functions, one obtains instead of (6) the following equations that became famous and are named Hartree – Fock equations:

$$-\frac{\Delta}{2}\varphi_j(x) - \frac{Z}{r}\varphi_j(x) + \sum_{k=1}^{N}\int \varphi_k^*(x') \frac{1}{|\vec{r}'-\vec{r}|}[\varphi_k(x')\varphi_j(x) - \varphi_k(x)\varphi_j(x')]d\vec{r}' \equiv$$
$$\equiv -\frac{\Delta}{2}\varphi_j(x) + U_H(r)\varphi_j(x) - \sum_{k=1}^{N}\int \varphi_k^*(x')\frac{1}{|\vec{r}'-\vec{r}|}\varphi_k(x)\varphi_j(x')d\vec{r}' \equiv \hat{H}_{HF}\varphi_j(x) = E_j\varphi_j(x)$$
, (8)

It is seen from (8) that the term $k = j$ does not contribute and thus the self-action is eliminated. At large distances the effective potential that is a combination of $U_H(r)$ and the exchange term in (8) at large distances decreases as $(-Z + N - 1)/r$ thus supporting even for neutral atom an infinite number of discrete levels, the wave functions of which as above being determined by (8) with $j$ denoting a level above N already occupied.

The contribution of the second term in the integrand cannot be presented as an action of some normal potential $W(r)$ upon $\varphi_j(x)$. On the contrary, the action described by this term is non-local, connecting points $x$ and $x'$, over which the integration is performed.

## 4. Extra zeroes and asymptotic of the wave function

It is well known (see e.g. [4]) that each discrete level in an attractive spherically – symmetric potential is characterized by the following set of quantum numbers: principal $n$, radial $n_r = n - l - 1$, angular momentum $l$ and its projection $m$, and spin projection $s$. It is known that the number of zeroes (or nodes) of the wave function $\varphi_{nlms}(x)$ is equal to $n_r$. The wave function of the lowest in energy state with $n = 1$, 1s, has a no zeroes.

We will show here that solutions of HF equations have extra zeroes, even for the lowest 1s level, if states with higher principal quantum numbers are occupied. The number of zeroes is not determined by the radial quantum number $n_r$ of the considered level, but mainly by $n_r$ of the outermost particle. The general proof of this statement one can find in [6]. It appeared that the extra zeroes are located at big distances. Having this in mind, let us consider the asymptotic of the one-particle HF wave function. Consider for simplicity a two-level "atom", with one inner $i$ and the other – outer $o$ and consider the equation for the inner state with wave-function $\varphi_i(x)$ in order to see how it is modified by the exchange with the outer electron. Instead of (8) we obtain

$$-\frac{\Delta}{2}\varphi_i(\vec{r}) - U_H(r)\varphi_i(\vec{r}) - \int \varphi_o^*(\vec{r}') \frac{1}{|\vec{r}' - \vec{r}|} \varphi_i(\vec{r}') \varphi_o(\vec{r}) d\vec{r}' = -|E_i|\varphi_i(\vec{r}), \tag{9}$$

where $U_H(r)\varphi_i(\vec{r})$ is of the state $i$ radius.

At large distances the term with the integral $\Re(r)$ behaves as

$$\Re(r)|_{r\to\infty} = \int \varphi_o^*(\vec{r}') \frac{1}{|\vec{r}' - \vec{r}|} \varphi_i(\vec{r}') \varphi_o(\vec{r}) d\vec{r}'\bigg|_{r\to\infty} \simeq \frac{1}{r^2} \varphi_o(\vec{r}) \int \varphi_o^*(\vec{r}')(\vec{r}'\vec{n})\varphi_i(\vec{r}') d\vec{r}'$$
$$\equiv \frac{C_n}{r^2} \varphi_o(\vec{r}), \tag{10}$$

where $\vec{n}$ is the unit vector in the direction $\vec{r}$ and index $n$ stands for the principal quantum number of outer electron.

It is evident from (10) that located inside "atom" $\varphi_i(\vec{r})$ acquires an admixture of $\varphi_i(\vec{r})$ with much bigger radius of the outer electron. The wave function of an outer electron has more zeroes than has the inner if exchange term is not taken into account. Let us consider as $i$ a 1s state $\varphi_{1s}(\vec{r}) = \alpha^{3/2} e^{-\alpha r}$, where $\alpha = \sqrt{2|E_{1s}|}$. As it follows from (10), it can be mixed with only $p$-states. Assume for simplicity that the wave functions are close to hydrogenic functions. Their asymptotic are $\varphi_{np}(\vec{r}) \simeq \beta_n^{3/2}(\beta_n r)^{n-1} e^{-\beta_n r}$. Substituting a superposition of asymptotes into (9), we obtain using (10) the following expression for asymptotic of $\varphi_i(\vec{r})$:

$$\varphi_i(\vec{r})|_{r\to\infty} \simeq \alpha^{3/2} e^{-\alpha r} - \frac{\beta_n^{3/2} C_n}{(\alpha r)^2}(\beta_n r)^{n-1} e^{-\beta_n r}, \tag{11}$$

If $\alpha$ is considerably bigger than $\beta$, i.e. if the energy levels are well separated, the first term in (11) can be neglected thus leading to

$$\varphi_i(\vec{r})|_{r\to\infty} \simeq -\frac{\beta_n^{3/2} C}{(\alpha r)^2}(\beta_n r)^{n-1} e^{-\beta_n r}, \tag{12}$$

thus completely modifying the asymptotic as compared to the case when exchange is neglected.
We see that the asymptotic of any one-electron HF occupied state wave function $\varphi_j(x) \sim \exp(-\sqrt{2E_j}r)$ is determined not by the state's binding energy $|E_j|$ but can be much bigger, $\sim \exp(-\sqrt{2E_{min}}r)$, where $E_{min}$ is the energy of the outermost particle. If there are several outer levels, the effect of exchange is by the following expression

$$\varphi_i(\vec{r})|_{r\to\infty} \simeq -\sum_{\text{All outer } n} \frac{\beta_n^{3/2} C}{(\alpha r)^2}(\beta_n r)^{n-1} e^{-\beta_n r}, \tag{13}$$

that for $N_o$ outer electrons enhances the exchange influence by this same factor. The role of exchange contribution can be achieved by exciting the outer electrons to states with smaller than minimal in the atomic ground state energies.

The alteration of the one-particle wave function could profoundly increase the probability of ionization of the inner levels by a strong laser fields.

As an example, we present in Fig. 1 the 1s wave function Hartree and HF wave functions. It is seen that exchange that add in accord with (10) an admixture of *p*- function acquire a zero at $r = 1.176$ at. un. If exchange I neglected, zeroes have *p*- functions only with $n > 2$. For Ar case it means that only admixture of 3*p* – state adds a zero to 1s.

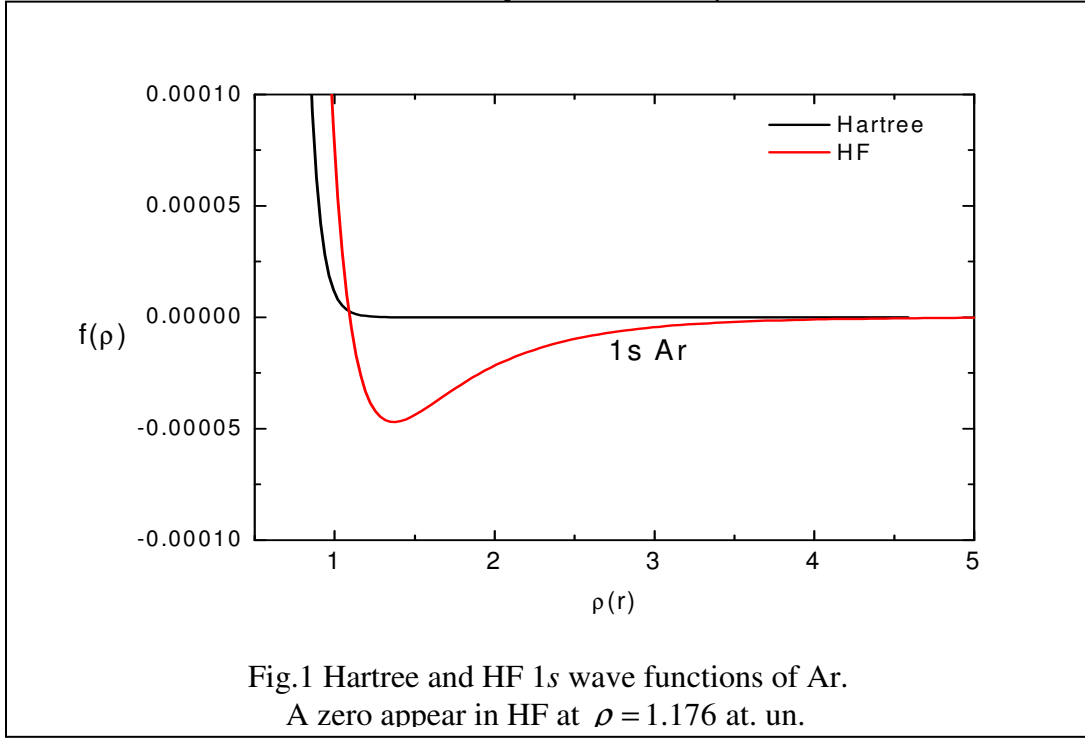

Fig.1 Hartree and HF 1*s* wave functions of Ar.
A zero appear in HF at $\rho = 1.176$ at. un.

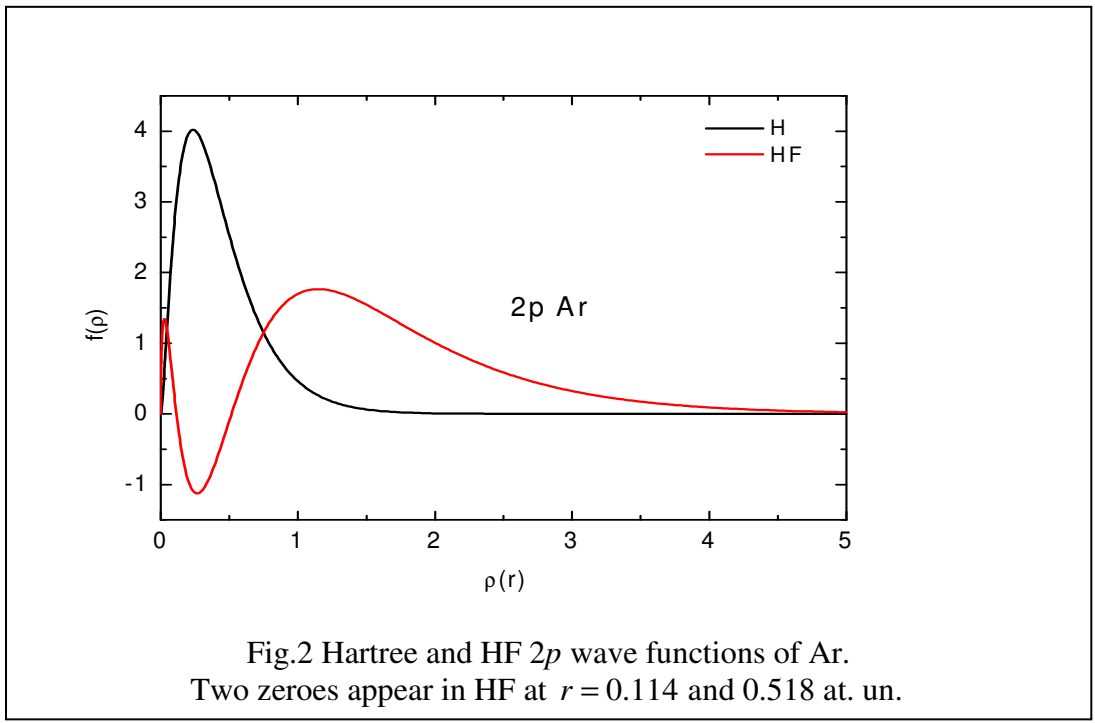

Fig.2 Hartree and HF 2*p* wave functions of Ar.
Two zeroes appear in HF at $r = 0.114$ and $0.518$ at. un.

that is confirmed by Fig.1. Note that $2p$ functions also do not have zeroes. If their exchange with $3s$ is taken into account, two extra zeroes appear in $\varphi_{2p}(r)$ as is demonstrated in Fig. 2.

The following notations are used in Fig. 1 and 2: $\rho = 2.79r + \ln r$ and $f_{nl}(\rho) = \sqrt{r(2.79r+1)}\varphi_{nl}(r)$. Substitution of $r$ by $\rho$ permits and choice of the numeric coefficient in front of $r$ permits more conveniently present the Hartree or HF wave functions using equidistant scale in $\rho$.

It is essential that inclusion of exchange increases the asymptotic value of the wave function by many orders of magnitude. In Ar the 1s wave-function at $\rho = 4$ at. un. increases by the factor $10^{17}$, while for 2p this factor is $10^5$. With further increase of $r$ corresponding factors increase even stronger.

## 5. Effect on multi-photon ionization

We will show here how long-tail corrections that appear in inner one-electron wave functions due to exchange with outer electrons modify the probability of their elimination from an atom by a strong electric field, of which a concrete example can serve a high intensity (about $10^{18-20}$ and higher Watts/cm$^2$) and low frequency ($\omega \ll I$) laser beam. The combination of static external and atomic field is depicted in Fig. 3. Let us consider for simplicity a two-level atom considered in the previous section. Here we will follow [7].

The probability to be ionized by the static field for electrons $i$ and $o$ is determined by the probability to find corresponding electrons at points $r_i$ and $r_o$. As is well known, this probability is given by square modulus of the corresponding wave functions at points $r_i$ and $r_o$. Assuming that these points correspond already to the asymptotic region for the wave function, we receives in the one-electron approximation for $i$ and $o$ electrons, respectively

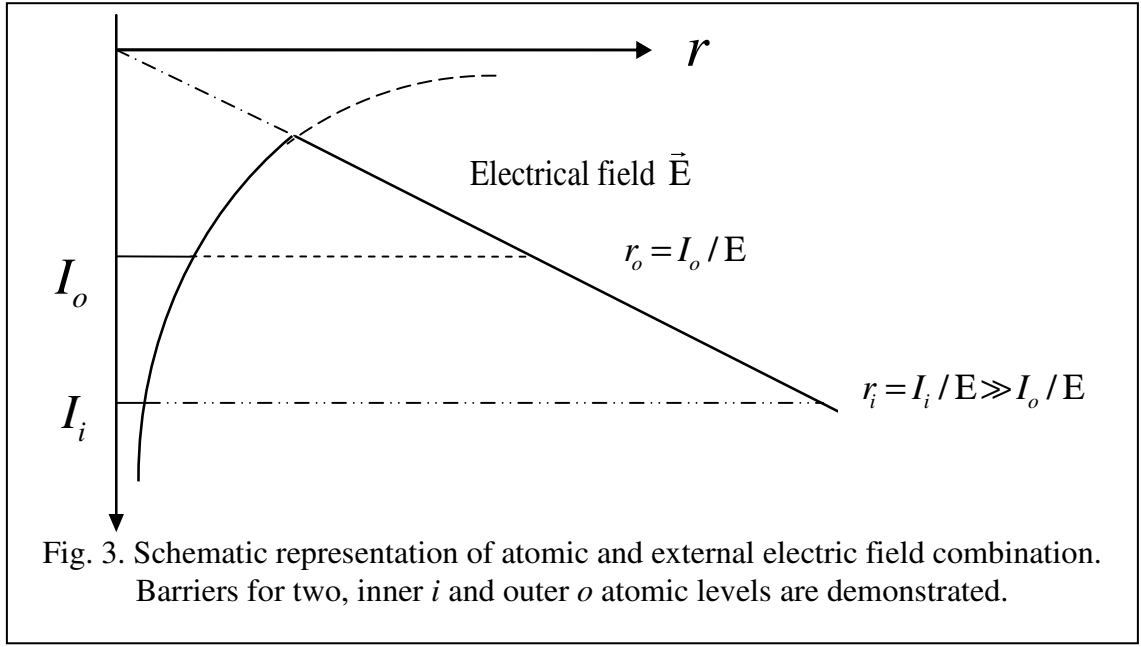

Fig. 3. Schematic representation of atomic and external electric field combination. Barriers for two, inner $i$ and outer $o$ atomic levels are demonstrated.

$$\begin{aligned}"i" \quad &|\varphi_i(r_i)|^2 \sim \alpha^3(\alpha r_i)^{2(n_i-1)}e^{-2\alpha r_i} \approx I_i^{n_i+1/2}\left(\frac{I_i}{E}\right)^{2(n_i-1)}e^{-2\sqrt{2I_i}\frac{I_i}{E}} \\ "o" \quad &|\varphi_o(r_o)|^2 \sim \beta_n^3(\beta_n r_i)^{2(n-1)}e^{-2\beta_n r_o} \approx I_o^{n+1/2}\left(\frac{I_o}{E}\right)^{2(n-1)}e^{-2\sqrt{2I_o}\frac{I_o}{E}}\end{aligned} \quad (14)$$

Let us now introduce inter-electron Coulomb interaction and exchange. According to consideration in the previous section, the wave function of inner electron is given by (11) that lead to another decay probability than (14), namely

$$"i" \quad |\varphi_{i,ex}(r_i)|^2 \approx \left| \alpha^{3/2} e^{-\alpha r_i} - \frac{\beta_n^{3/2} C_n}{(\alpha r_i)^2} (\beta_n r_i)^{n-1} e^{-\beta_n r_i} \right|^2 \quad (15)$$

For deep levels the contribution of the first term is negligible, so that the penetration of the electron out of the atom is given by expression

$$|\varphi_{i,ex}(r_i)|^2 \approx \frac{\beta_n^3 C_n}{(\alpha r_i)^4} (\beta_n r_i)^{2n-2} e^{-2\beta_n r_i}. \quad (16)$$

The enhancement factor $\eta$ due to inclusion of the Fock term into the one-electron wave function of the inner electron is determined by the ratio of (16) to the expression in the first line in (14)

$$\eta \equiv \frac{|\varphi_{i,ex}(r_i)|^2}{|\varphi_i(r_i)|^2} = C_n^2 (\beta_n r_i)^{2(n-n_i)} e^{2(\alpha-\beta_n)r_i} \approx C_n^2 (\sqrt{2I_o} I_i / \mathrm{E})^{2(n-n_i)} \exp(2\sqrt{2I_i} I_i / \mathrm{E}) \quad (17)$$

It is remarkable that if there are $N_o$ outer electrons, the factor $\eta$ in accord with (13) acquire an additional enhancement factor $N_o^2$. To illustrate the possible size on the enhancement factor $\eta$, let us consider a numerical example, in which the inner electron binding energy $I_i$ is five atomic units, while the outer electron binding energy $I_o$ is half atomic unit and the external field intensity E is one unit[4]. Then the factor $\eta$ is of the order of 5.64x10$^{13}$, while for the same field and levels energy one and ten atomic units, respectively one have 7.86x10$^{38}$! These tremendously big numbers are a consequence of extremely small probability to eliminate an inner electron without exchange with the outer. Therefore it is more interesting and instructive to compare the ratio $\tau$ of inner to outer electron ionization probabilities when the exchange between outer and inner electrons is taken into account. This ratio is given by the following expression

$$\tau = \frac{|\varphi_{i,ex}(r_i)|^2}{|\varphi_o(r_o)|^2} \approx \frac{\beta_n^3 C_n^2}{(\alpha r_i)^4} (\beta_n r_i / r_o)^{2n-2} e^{-2\beta_n (r_i - r_n)} \approx$$
$$\approx \frac{\beta_n^3 C^2}{(\alpha r_i)^4} (\beta_n r_i / r_n)^{2n-2} e^{-2\sqrt{2I_n} I_i / \mathrm{E}} \sim \mathrm{E}^4 \exp(-2\sqrt{2I_n} I_i / \mathrm{E}). \quad (18)$$

For the considered examples the energies of two levels, it is obtained for $\tau \approx 4.49 \times 10^{-5}$ and 5.01x10$^{-13}$. For the first case the ratio is not too small.

Qualitatively, it looks like the exchange admixture of outer electron literally "drags out" the inner electron off the ionized atom.

As it was mentioned before, if it is $N_o$ outer electrons, for which the coefficient $C$ is non-zero, this ratio is increased by additional factor $N_o^2$. It seems that this dependence was really observed in a number of investigations (see e.g. [8]) observed in studies of multiple photoionization of noble-gas clusters by high intensity laser beam. In this studies a prominent amount of photons with energies of several hundreds eV were found signaling the possibility that vacancies in inner shells were generated during laser-cluster interaction. The intensity of such processes in clusters could be a direct consequence of presence of very many outer electrons in clusters, contrary to the case of isolated atoms.

---
[4] That corresponds to the field intensity 10$^{16}$ W/cm$^2$.

Note that exchange effects could be strongly enhanced even if the target atom exists in an exited state for a relatively short time. Therefore presence of strong atomic resonances, e.g. Giant, at laser frequency can enhance the multiple ionization probability considerably. Perhaps this is the reason why in photoionization by free-electron laser an abundance of multiply charged ions, with degree of ionization up to twenty-one were found [9].

It was demonstrated recently both numerically and analytically in [10, 11] that the Fock exchange leads in fact to non-exponential instead of exponential barrier penetration probability. This is seen qualitatively already from [11]: if $\beta_0 r_i \sim 1$ the second term presents power decrease of the barrier penetration probability.

## 6. Violation and restoration of the Gauge invariance

The operator of photon-electron interaction in non-relativistic approximation is given by (4), that is called in atomic physics "length form". In HF it is not equivalent to the so called "velocity form" $V_{pe} = i\vec{A}(\vec{r},t)\vec{\nabla}/c$. Indeed, the velocity $\vec{v}$ in HF is not equal to $i\vec{\nabla}$. Instead, according to (8), one has in HF approximation

$$\vec{v}_{HF} \equiv \partial \vec{r}/\partial t = i[\hat{H}_{HF}, \vec{r}] = i\hat{\vec{\nabla}} - \sum_{k=1}^{N} \int \varphi_k^*(x') \frac{1}{|\vec{r}' - \vec{r}|} \varphi_k(x)(\vec{r}' - \vec{r}) \, dx' \neq i\hat{\vec{\nabla}}. \quad (19)$$

The difference between $\vec{v}_{HF}$ and $i\hat{\vec{\nabla}}$ is a consequence of the fact that non-locality is equivalent to velocity dependence of a non-local potential. Indeed, a non-local operator can be presented as an infinite sum of local velocity dependent terms:

$$V(\vec{r}, \vec{r}') = e^{(\vec{r}' - \vec{r})\vec{\nabla}} V(\vec{r}, \vec{r}), \quad (20)$$

where $\vec{\nabla}$ act upon the second variable in $V(\vec{r}, \vec{r})$. It is known that to include interaction with electromagnetic wave in a gauge invariant form it is necessary to substitute electron linear momentum operator in the following way $\hat{\vec{p}} \to \hat{\vec{p}} - A(\vec{r})/c$ or $\vec{\nabla} \to \vec{\nabla} + iA(\vec{r})/c$. Using (20) and the definition of Fock non-local potential $F(\vec{r}, \vec{r}')$ that follows directly from (8), the photon-electron interaction operator $V_{pe}$ instead of (4) can be presented in the following way [12]:

$$V_{pe} = i\frac{1}{c}\vec{A}(\vec{r},t)\vec{\nabla} + \int d\vec{r}' \left\{ \exp\left[i(\vec{r} - \vec{r}')\vec{A}(\vec{r},t)\right] - 1 \right\} F(\vec{r}, \vec{r}'). \quad (21)$$

Note that (21) is obtained if for $\vec{A}(\vec{r},t)$ A the Coulomb gauge is chosen $\text{div}\vec{A}(\vec{r},t) = 0$. For precise atomic Hamiltonian (1) "length" and "velocity" forms in calculating the photoionization cross-section give the same result. However, the difference between corresponding values calculated in HF does not characterize the precision of the wave-function, but rather the HF field's non-locality. Concrete calculations [5] have demonstrated that this difference is quite big.

In HF the sun rule (5) is also violated. Instead, one has [12] the following relations

$$S_r = N + \langle 0 | [Z, [\hat{F}, Z]] | 0 \rangle = N + \Delta S_r > N,$$
$$S_\nabla = N - \langle 0 | [Z, [\hat{F}, Z]] | 0 \rangle - \sum_n \frac{2}{\omega_{n0}} \left| \langle 0 | [\hat{F}, Z] | n \rangle \right|^2. \quad (22)$$

Here $\hat{F} \equiv \int F(\vec{r}, \vec{r}') d\vec{r}'$ and the average is taken over HF function of the atom in the ground state. In statistical approximation one can easily estimate $\Delta S_r$ as $\Delta S_r \sim N^{2/3}$. In the expression for $\Delta S_\nabla$ $\omega_{n0}$ is the one-electron excitation HF energy.

To restore gauge-invariance violated in HF approximation, one has to go beyond it, namely to the Random Phase Approximation with Exchange [5], or to the Time-Dependent HF [13]. Note that the width of the Giant resonance in photo-absorption cross-section of the system is determined mainly by non-locality.

## 7. Green's function

The Green function $G_E(\vec{r}, \vec{r}')$ of HF, as of any other one-particle Schrödinger equation, is determined by the following equation

$$[\hat{H}_{HF}(\vec{r}) - E] G_E(\vec{r}, \vec{r}') = \delta(\vec{r}' - \vec{r}). \quad (23)$$

It easy to show that $G_E(\vec{r}, \vec{r}')$ can be presented as

$$G_E(\vec{r}, \vec{r}') = \sum_k \frac{\varphi_k^*(\vec{r}) \varphi_k(\vec{r}')}{E_k - E}, \quad (24)$$

where summation includes also integration over the continuous spectrum, $\varphi_k(\vec{r})$ are regular at $r = 0$ solutions of the HF equations $[\hat{H}_{HF}(\vec{r}) - E_k] \varphi_k(\vec{r}) = 0$. These solutions obey the completeness relation $\sum_k \varphi_k^*(\vec{r}) \varphi_k(\vec{r}') = \delta(\vec{r} - \vec{r}')$.

Along with regular solutions $\varphi_k(\vec{r})$, the Schrödinger equation has irregular at $r = 0$ solutions $\chi_k(\vec{r})$. It is known that the one-particle Green's function can be expressed not only as (24), but also via product of regular and irregular solutions of the one-particle Schrödinger equation

$$G_E(\vec{r}, \vec{r}') = \varphi_p(\vec{r}_<) \chi_p(\vec{r}_>), \quad (25)$$

where $p$ denotes a state with energy $E$, $\vec{r}_{>(<)}$ denotes a radius that is bigger (smaller) among $\vec{r}, \vec{r}'$ by modulus.

Substituting (25) into equation (23), one can easily see that due to non-locality of the Fock term, (25) in HF is not a correct expression for the Green's function. It is a pity, since

(25) is convenient in performing calculations, permitting substitute summation over infinite set of states by solving Schrödinger equation for $\chi_p(\vec{r})$.

## 8. Levinson's theorem

Almost in any book on quantum mechanics or quantum scattering one can find Levinson's theorem (see e.g. [4]). It reads that if one determines scattering phases $\delta_l(E)$ of any partial wave in such a way that they became zero at infinite scattering energy[5], the phase at zero energy $\delta_l(0)$ is given by relation

$$\delta_l(E) = n_l \pi, \qquad (26)$$

where $n_l$ is the number of bound states of the incoming particle in the potential it is scattered by.

Direct calculations [14] and analytic evaluation permitted to find that for HF equation due to effect of Fock term (26) transforms into

$$\delta_l(E) = (n_l + n_{lo})\pi, \qquad (26)$$

where $n_{lo}$ is the number of *occupied* states with angular momentum $l$ in the target atom itself. This is very substantial change since for neutral atoms in HF there are at most one-two bound states. As to the occupied levels, atom Ar has three *s*-states and two *p*-states that gives $n_{so} = 3$ and $n_{1o} = 2$.

## 9. Non-uniqueness of HF solutions

Non-linearity leads to non-uniqueness of HF solutions for the same energy. This is demonstrated in the frame of two models for inter-particle interaction $V(\mathbf{r}_i, \mathbf{r}_k)$. For the first time it was done in [15].

Since in this particular case the main role is played by Hartree term, to simplify consideration we will limit ourselves with Hartree equations only. In this case it looks as [see (6)]

$$\left[-\frac{\Delta}{2} - \frac{Z}{r} + \sum_{i=1}^{N} \int d\mathbf{r}' |\varphi_i(\mathbf{r}')|^2 V(\mathbf{r}',\mathbf{r})\right] \varphi_k(\mathbf{r}') = E_k \varphi_k(\mathbf{r}) \qquad (27)$$

The total Hartree energy $E$ of the system that is of interest for our consideration is defined by

$$E = \sum_{k=1}^{N} E_k - \frac{1}{2} \sum_{i \neq k=1}^{N} V_{ik,ik}. \qquad (28)$$

We now choose a model potential $V(r_1, r_2)$ such that Hartree equations can be solved analytically. Consider a potential of the form

---
[5] No effect of scattering at very high collision energy that is quite natural.

$$V(\mathbf{r}_1, \mathbf{r}_2) = \frac{\alpha}{r_1 \cdot r_2} \tag{29}$$

instead of the usual Coulomb potential in (6).

In (29) $\alpha$ is a constant and considered positive. We assume that the potential (29) describes the interaction of the electrons that fill a subshell with quantum numbers $n$ and $l$ in the atom. Let half of these electrons describe by the wave function $\mathbf{r}\,\varphi_1(\mathbf{r})$, and the other half by $\varphi_2(\mathbf{r})$. Then from (27) and (29) we have

$$\left(-\frac{\Delta}{2} - \frac{Z_{ef}^{(p)}}{r}\right)\varphi_p(\mathbf{r}) = E_p \varphi_p(\mathbf{r}), \quad p=1,2 \tag{30}$$

$$Z_{ef}^{(p)} = Z - \alpha \sum_{i \neq k=1}^{N} \int d\mathbf{r}' |\varphi_i(\mathbf{r}')|^2 \frac{1}{r'}. \tag{31}$$

Ordinary wave functions for the Coulomb field of the charge $Z_{ef}^{(p)}$ satisfy the equation (30). We calculate the integral in (31) to find a system of equations for $Z_{ef}^{(p)}$

$$Z_{ef}^{(p)} = Z - \frac{\alpha}{n^2}\left[2l Z_{ef}^{(p)} + (2l+1) Z_{ef}^{(q)}\right]. \tag{32}$$

Here $p, q = 1, 2$ and $p \neq q$. These equations have the obvious solution

$$Z_{ef}^{(1)} = Z_{ef}^{(2)} = Z\left[1 + \frac{\alpha}{n^2}(4l+1)\right]^{-1}, \tag{33}$$

which will be called normal solution. If the constant $\alpha$ satisfies the condition $\alpha = n^2$, then the system (32) degenerates into a single equation

$$\sum_{p=1,2}(2l+1)Z_{ef}^{(p)} = Z, \tag{34}$$

that has an infinite set of solutions, for which $Z_{ef}^{(1)} \neq Z_{ef}^{(2)}$, $Z_{ef}^{(1)}, Z_{ef}^{(2)} > 0$. These solutions we will call singular. The radial wave functions and one-particle energies of all the electrons in subshell $nl$ are identical for the normal solution. For a singular solution, the one-particle energies and the wave functions $\varphi_1(\mathbf{r})$ and $\varphi_2(\mathbf{r})$ are different. From (28) we find the total energy of the system,

$$E = -\frac{Z^2 N}{2n^2}\left[1 + \frac{\alpha}{n^2(N-1)}\right]^{-1}, \tag{35}$$

which coincides for the normal solution and all the singular ones. In deriving (35) it was taken into account that $N = 4l + 2$.

Note that at the point of instability, i.e. at $\alpha = n^2$ the total energy $E$ reduces to that of a single particle energy in the Coulomb field of the nucleus with the charge Z:

$$E = -Z^2 / 2n^2, \qquad (36)$$

and is independent upon N.

With increase of $\alpha$ from 0 to $n^2$ the mean radius R of the system under consideration is growing smoothly, but at $\alpha = n^2$ becomes non-determined being able to acquire any value, even $R \to \infty$. Sets of solutions with $Z^{(1)} \neq Z^{(2)}$ have the same principal quantum number, namely $n$. So contrary to the case of linear Schrödinger equation, $n$ is unable to determine uniquely the solution of Hartree nonlinear equation.

The singular solution occurs at a definite strength of inter-particle interaction. One solution may go over to the other in a continuous fashion as the parameter approaches the critical value, the energy of the system being preserved, while the size of it can be arbitrary.

Note, that at $\alpha = n^2$ the energy required to eliminate a particle according to (36), is equal to zero, while the Hartree energy of a particle is nonzero and is almost arbitrary. This difference is a direct manifestation of essential rearrangement of Hartree states after eliminating a particle off the considered system.

The existence of singular solutions seems to be a reasonable general feature of Hartree equations and is independent of the particular form of the particle interaction. In the above model, the singular solution occurred without relation to any special feature of the shape of the potential well, but solutions arise only from the non-linearity of the equations.

Indeed, we can consider a Hamiltonian, different from (1) [16]:

$$\hat{H} = -\sum_{i=1}^{N} \frac{\Delta_i}{2} - \sum_{i=1}^{N} \frac{1}{2} \omega^2 r_i^2 + \frac{1}{2} \sum_{i \neq k=1}^{N} V(\mathbf{r}_i, \mathbf{r}_k), \qquad (37)$$

which describes $N$ particles moving in an oscillator potential and interacting each other via

$$V(\mathbf{r}_1, \mathbf{r}_2) = -\beta r_1^2 \cdot r_2^2. \qquad (38)$$

Then instead of (6) and (7) one gets

$$\left(-\frac{\Delta}{2} - \frac{1}{2}(\omega_{ef}^{(k)})^2 r^2\right)\varphi_k(\mathbf{r}) = E_k \varphi_k(\mathbf{r}), \qquad (39)$$

$$(\omega_{ef}^{(k)})^2 = \omega^2 - \beta \sum_{i \neq k=1}^{N} \int d\mathbf{r}' |\varphi_i(\mathbf{r}')|^2 r'^2. \qquad (40)$$

The solutions of (39) are known and the integrals in (40) may be calculated in a closed form. Separating all the $N$ particles into two groups as it was done above in deriving (30) and (31), a system of two equations is obtained, which may be semi-schematically presented as

$$(\omega_{ef}^{(p)})^2 = \omega^2 - \tilde{\beta}(\omega_{ef}^{(q)})^2, \; p \neq q = 1, 2, \qquad (41)$$

where $\beta$ differs from $\tilde{\beta}$ by a factor, dependent upon the principal quantum number $n$ of the state, occupied by half $N$ particles as well as upon the number of particles in the group 1 and 2, $N_1$ and $N_2$ respectively. They are related to $N$ by equation $N=N_1+N_2$. The system of equations (41) has an obvious solution

$$(\omega_{ef}^{(1)})^2 = (\omega_{ef}^{(2)})^2 = \omega^2(1+\tilde{\beta})^{-1}, \qquad (42)$$

which is a normal one. However for $\tilde{\beta}=1$ the system (41) becomes degenerate and have an infinite number of solutions, the singular ones, satisfying the equation

$$(\omega_{ef}^{(1)})^2 + (\omega_{ef}^{(2)})^2 = \omega^2. \qquad (43)$$

As in the first example, the corresponding states can be arbitrary stretched in space by choosing one of the frequencies, $\omega_{ef}^{(1)}$ or $\omega_{ef}^{(2)}$ sufficiently small.

Another, rather interesting example of a Coulomb field Hamiltonian with two-body interactions $\gamma/r_1^2 \cdot r_2^2$ was also discussed in [15]. It was demonstrated there that starting from some value of $\gamma$ two instead of one solution appear.

Traces of the non-uniqueness described above with two instead of one solution were observed in numerical calculation of Hartree-Fock states for some atoms [17] where the role of nonlinearity was not clarified [18].

It is natural to examine the relationship between these solutions and the exact solution to the Schrödinger equation with an inter-particle interaction potential (29) or (38). Before answering this question, it is necessary to examine the stability of the solutions, which can be easily done for small disturbances. Normally, only one set of solutions will be stable against the transition of the system to the state, described by the other set (or sets).

The stability of both normal and singular solutions against small perturbations can be examined by solving the Random-Phase Approximation (RPA) equation, derived in [13]. The corresponding equations determine whether an excitation with zero energy exists in the considered system, which means a possibility of a transition into another state with the same energy. For the choice of interparticle interaction (4) RPA equations are simplified and reduced to

$$1 + \alpha\chi = 0, \qquad (44)$$

where $\chi$ is given by

$$\chi = \sum_j \left|\langle j|r^{-1}|i\rangle\right|^2 (E_j - E_i). \qquad (45)$$

In (45) summation is performed over all vacant states $j$, which energy is given by solution of (30). In (45) the following notation is used $\langle j|r^{-1}|i\rangle \equiv \int \varphi_j^*(\mathbf{r})r^{-1}\varphi_i(\mathbf{r})d\mathbf{r}$.

Equation (44) determines the $\alpha$ value, at which considered Hartree state $\varphi_i(\mathbf{r})$ becomes unstable. For interparticle interaction (38) and Hamiltonian (37) one has an instability equation similar to (44) with $\beta$ instead of $\alpha$ and matrix elements of $r^2$ instead of $r^{-1}$ in $\chi$.

It is possible in principle to have a situation where the normal solution and the singular one are both stable. Then these should describe two isomeric states whose energies are quite close, but whose charge distributions or sizes are very different. These states have identical quantum numbers, so a radiative transition between them is forbidden. If one of them is a ground state, the other will be metastable. It is desirable to search for such states. For the atomic case, they may be excited in heavy ion collisions. The photoionization cross-sections of the ground and metastable states will also differ considerably, which might be a way to discover them.

## 10. Concluding remarks

In this paper we have presented a number of consequences that are results of non-linearity and non-locality of the Hartree-Fock equations. As an example, we concentrate on atoms. However, HF equations are used far beyond pure physics of atoms. This is why the result presented in this paper may be essential in other domains, where the HF managed to penetrate.

Very often HF results are compared to that obtained in the frame of LDA – Local Density Approximation [19]. It is necessary to have in mind, however, that LDA by definition lacks non-locality. This is why in some respect the results of calculations using both these approaches can differ considerably.

As appropriate objects for HF equations, however much more difficult for calculations, are atoms imbedded in condensed matter objects, clusters or fullerenes. However, since they have much more outer electrons than an isolated atom, the exchange and density effects have to be in these objects stronger.

One can expect traces of exchange effects in atomic collisions, while temporarily strongly exchanging objects are formed.

Of interest are also effects of exchange and non-linearity also in two and even one-dimensional HF equations.

HF equations were studied and applied mainly to multi-fermion systems. For bosons the exchange correction does not eliminate self-action, leading instead to doubling of its effect. However the whole concept of the role of non-locality and non-linearity could be of interest there also.

## 11. Acknowledgements

I am grateful to Prof. L. V. Chernysheva for supplying me with Ar wave functions.